# Accelerating Light with Metasurfaces

M. Henstridge,[1,2*] C. Pfeiffer,[1,3] D. Wang,[1,4] , A. Boltasseva[1,4], V. M. Shalaev[1,4],

A. Grbic[1,3], and R. Merlin[1,2]

[1]*Center for Photonics and Multi-Scale Nanomaterials, University of Michigan,*

*Ann Arbor, 48109, USA*

[2]*Department of Physics, University of Michigan, Ann Arbor, MI 48109-1040, USA*

[3]*Department of Electrical Engineering and Computer Science,*

*University of Michigan, Ann Arbor, MI 48109-2122, USA*

[4]*School of Electrical and Computer Engineering and Birck Nanotechnology Center,*

*Purdue University, West Lafayette, IN 47907, USA*

*\*corresponding author: mhenst@umich.edu*

It has been recently shown that especially engineered light beams have the remarkable ability to propagate along curved trajectories in vacuum. Current methods for generating accelerating beams use phase modulators and lenses leading to length scales on the order of tens of centimeters or larger. This poses constraints and severely limits their applicability inside materials. Here, we accelerate light inside glass using a metasurface consisting of plasmonic nanoantennas. Highly-bending beams with radii of curvature on the order of a hundred microns were generated, and the imaged intensities agree well with theory. Our approach for generating accelerating beams allows for their integration into on-chip photonic systems.

Non-diffracting solutions to Maxwell's equations other than plane waves, such as Bessel [1,2] and Mathieu [3] beams, which propagate indefinitely along linear trajectories without transverse distortions, have been known for a long time. There is a lesser-known and particularly unique group of diffraction-free solutions for which the intensity maximum follows a curved trajectory. In contrast to beams that accelerate due to refractive index gradients, such as those that form mirages [4] or those inside a gradient-index (GRIN) system [5], these solutions to Maxwell's equations propagate along curved trajectories in vacuum. As with all diffraction-free solutions, accelerating beams have both infinite energy and spatial extent so that the approximate beams that are realized in practice preserve their shape over a finite distance before diffracting.

Accelerating force-free fields were first reported by Berry and Balazs [6] who showed that Airy function solutions to a one-dimensional Schrodinger's equation propagate with constant acceleration. Later, it was shown that Airy solutions exist also for the two-dimensional Helmholtz equation (HE) under the paraxial approximation [7]. Their finite-energy realization [8] was the first demonstration of accelerating light. Subsequent investigations revealed exact, non-diffractive, accelerating solutions to the HE involving Bessel and Mathieu functions that propagate along closed elliptical trajectories [9,10]. Practical versions of such beams [10,11] and also others from caustics theory [12-14] have the unusual ability to bend at angles well-beyond those of the paraxial-limited Airy beams. These accelerating, non-diffractive solutions should be distinguished from the non-diffractive and non-accelerating Bessel and Mathieu beams, which propagate along straight lines [1,3]. Applications of accelerating beams to microparticle manipulation [15,16], micromachining [17], optical circuits [18] and nonlinear optics [19-24] have been reported, and we note that the concept of accelerating waves has been extended to surface plasmon polaritons [25-30].

The most common method for generating accelerating beams relies on spatial light modulators (SLMs) [8,11-18,22,31], although other techniques using phase masks [19,20] and lens aberrations [24,32] have been demonstrated. SLMs, in particular, are severely limited in their ability to bend light at steep angles because of the large size of their pixels. For light of 800 nm wavelength at normal incidence, the first-order diffraction angles of commercial SLMs range from $\approx 1°$ to $\approx 15°$ at best [33]. In order to generate beams that bend beyond these angles, additional lenses must be used to focus the diffracted light. At a minimum, this extends the size of the setup to many centimeters, preventing the use of accelerating beams for sub-centimeter-sized and on-chip systems. Phase masks have a similar constraint, as the corresponding accelerated-beam setups require additional optics [19,20]. Another limitation, common to all of the aforementioned approaches, arises when an application requires accelerating light inside a material, since Snell's law constrains the ability of light incident from free space to bend at steep angles [34]. Standard water and oil immersion optics used to reduce index mismatch add considerable complications to the setup and are less-effective for higher index materials. Thus, in addition to their large size, current setups become difficult to implement when attempting to accelerate beams in materials, especially in those with high values of the refractive index.

In this letter, we demonstrate a broadly applicable, planar optic approach for the acceleration of light inside a material using metasurfaces. The sub-wavelength size and spacing of the metasurface constituents allow for the bending of light at arbitrarily high angles in any medium, lifting the constraints of conventional optics and eliminating the limitations associated with the above-mentioned setups. Metasurfaces have been used to manipulate light for a number of applications, some of which include near-field focusing [35,36], nonlinear light generation [37-

39], ultra-thin holograms [40,41], invisibility cloaks [42], and negative refraction [43,44]. Until now, the metasurface approach has not, to our knowledge, been used for the acceleration of light.

Our metasurfaces consist of gold V-shaped nanoantennas [43,44]. In contrast to dielectric metasurfaces, which often require harsh chemical etches [45] or time-consuming and expensive atomic layer deposition processes [46], the fabrication process for the nanoantennas is easily optimized and can be used on a wide-variety of substrates, making large-scale fabrication for on-chip applications considerably more feasible. Furthermore, dielectric metasurfaces typically have larger unit cell sizes [45,46] which can impose limitations when generating beams with high numerical apertures inside materials.

The nanoantennas are characterized by their arm length $L$ and the angle between the arms, $\Delta$; see Fig. 1(a). These antennas exhibit two so-called symmetric and antisymmetric resonances at, respectively, $\lambda_{eff} \approx 2L$ and $\lambda_{eff} \approx 4L$, where $\lambda_{eff}$ is an effective wavelength which depends on the vacuum wavelength, plasma frequency, antenna cross section, and surrounding medium [47].

Excitation of the symmetric or antisymmetric resonance alone occurs when the incident field is polarized respectively along the principal S or AS axis, which are perpendicular to each other [see Fig 1(a)]. In both cases, the scattered field is polarized along the same direction as that of the incident field, and the scattering phase shift depends on both $L$ and $\Delta$. Because of their large bandwidth [43,44], both resonances can be simultaneously excited with monochromatic light if the excitation field is polarized along directions other than those of the principal axes. The resulting scattered field has polarization components both perpendicular (cross-polarized) and parallel (co-polarized) to the incident beam. We designed a set of antennas that operates for incident fields of wavelength 800 nm polarized at 45° with respect to both the principal axes. Central to our design is the fact that the phase shift for the cross-polarized configuration can be varied from 0 to $\pi$ by tuning *L* and Δ, and a total of 2π phase coverage can be achieved by rotating each antenna in the designed set by 90 degrees [43].

We simulated the scattered field of each antenna under the local periodicity approximation (LPA) [48], which assumes that the response of a single V-antenna is the same regardless of whether it is surrounded by antennas with identical or slightly differing parameters. The LPA assumption was tested by simulating the response of an array designed to refract normally incident light at an angle of 52°. We found that the amount of cross-polarized power transmitted in the desired direction was 40 times greater than for other directions.

The metasurfaces were configured to produce an accelerating beam inside a glass substrate upon illumination with light incident at an angle of 45° from the normal to the glass-vacuum interface; see Fig. 2. The nanoantennas scatter ~ 5% of the incident power into the cross-polarized accelerating beam whereas ~ 20% of the incident power is reflected. Approximately 50% of the incident power passes through the metasurface unperturbed and is refracted inside the glass according to Snell's law. As illustrated in Fig. 1(b), illumination at 45° allows for the spatial separation of the accelerating and unperturbed transmitted beams.

Although not pursued here, we note that Huygens metasurfaces [49,50] present an advantage over the V-antennas in that they allow control of the phase of both, cross-polarized and co-polarized scattered fields. Unlike the V-antenna arrays, which only provide control of the electric polarization, Huygens metasurfaces grant also manipulation of the magnetic response [49,50]. This leads in particular to a significant reduction in the intensity of the unperturbed transmitted and reflected co-polarized beams. However, the design required for control of the magnetic response at near-infrared wavelengths involves stacks of patterned metallic sheets [50], which adds complexity to the fabrication process.

The nanoantenna pattern was designed using back-propagation so that the electric field at the metasurface plane gives the proper field at a distance $z_0 = 300$ μm from the metasurface, when illuminated with light of frequency $\omega_0/2\pi = 3.75 \times 10^{14}$ Hz. We chose this field $E(z_0, x)$ to be the so-called half-Bessel beam [11]:

$$E(z_0, x) = \exp(\alpha x) J_\beta(-kx + \beta) \Theta\left(-x + \frac{\beta}{k}\right) \quad (1)$$

Here, $\alpha$ is a cut-off parameter, related to the truncation of $J_\beta(-kx + \beta)$, and $\beta = kR$, where $k = nk_0$ is the wavenumber inside the glass of index $n$, $k_0 = \omega_0/c$ is the wavenumber in vacuum, $R$ is the predetermined radius of curvature of the accelerating beam, and $\Theta$ is the Heaviside step function. We note that the full accelerating Bessel solution $J_\beta(k\sqrt{x^2 + z^2})\exp[i\beta \tan^{-1}(x/z)]$ cannot be realized in practice because it is not square-integrable, that is, it carries an infinite amount of energy. We used reverse propagation [35,36] to obtain the desired field at the metasurface plane, at $z = 0$:

$$E(0, x) = A(0, x)\exp[i\varphi(x)] = -2 \int E(z_0, x') \frac{\partial}{\partial z_0} G^-(z_0, x - x') dx' \quad (2)$$

where

$$G^-(z, x) = (i/4) H_0^{(2)}(k\sqrt{z^2 + x^2}) \quad (3)$$

is the backward-propagating Green's function for the 2D Helmholtz equation and $H_0^{(2)}$ is the 0th order Hankel function of the second kind. Our calculations show that the amplitude function $A(x)$ is approximately Gaussian across the metasurface plane. This allows for a design whereby the accelerating beam is generated solely by phase modulation of an incident Gaussian light source. To account for the angle of incidence of 45°, the phase shift at the metasurface plane must be $\varphi_m(x) = \varphi_r(x) + \varphi(x)$ where $\varphi_r(x) = k_0(\sqrt{2}/2)(x - \ell/2)$ and $\ell$ is the length of the metasurface along the $x$-axis. Based on these required phase shifts, the nanoantennas were arranged in a square lattice of spacing 220 nm, forming a discretized version of $\varphi_m(x)$ on the metasurface plane. An example of a metasurface-induced phase shift is shown in Fig. 1(c). Upon illumination, the accelerating beam at arbitrary distances from the metasurface is obtained by reversing Eq. (2) so that

$$E(z,x) = 2 \int E(0,x') \frac{\partial}{\partial z} G^+(z, x-x') dx' \quad (4)$$

where

$$G^+(z,x) = (i/4)\, H_0^{(1)}\left(k\sqrt{z^2 + x^2}\right) \quad (5)$$

is the forward propagating Green's function and $H_0^{(1)}$ is the 0$^{th}$ order Hankel function of the first kind.

We designed two metasurfaces that generate accelerating beams with radii of curvature of 400 and 100 µm, both with $\alpha = 0.01$ µm$^{-1}$. The nanoantennas were fabricated with standard electron-beam lithography on ITO-coated glass, followed by electron-beam evaporation of 3 nm Ti and 30 nm Au and lift-off processes. Optical microscope and scanning electron microscope (SEM) images of one of our metasurfaces are shown in Fig. 2(a).

The metasurfaces were illuminated with a mode-locked Ti: Sapphire oscillator which produced pulses of central wavelength 800 nm and bandwidth of 30 nm at a 78 MHz repetition rate. As discussed earlier, the $y$-polarized output from the laser was incident onto the metasurface at an angle of 45 degrees after being loosely focused with a lens, which generated an accelerating beam polarized in the $x$-$z$ plane; these axes are defined in Fig. 2(b). We used a standard setup to image the accelerating beam by collecting Rayleigh-scattered light from defects inside the glass. The polarization dependence of the Rayleigh cross-section helped reduce the signal of the $y$-polarized unperturbed transmission in the images. Fig. 3 shows a comparison between the experimental images of the accelerated beams and their theoretical intensity. The features of the experimental trajectories resolved by the imaging system agree very well with those predicted by simulations.

In conclusion, we accelerated light inside a material using a single planar optic. Our metasurfaces have an unrestricted ability to bend light, and simple tunings of their design can produce beams with radii of curvature ranging from several to thousands of wavelengths. This highly versatile method opens new opportunities for the integration of accelerating beams into on-chip photonic systems and for advanced studies of accelerating light in materials.

ACKNOWLEDGEMENTS

Work supported by the MRSEC program of the National Science Foundation (DMR-1120923) and the Air Force Office of Scientific Research MURI Grant (FA9550-14-1-0389). M. Henstridge thanks Dr. Pilar Herrera-Fierro, from the Lurie Nanofabrication Facility at the University of Michigan, for advice and training with the edge-polishing process. The electron beam lithography work used the EBPG 5200 system at the University of Notre Dame Nanofabrication Facility. Metallization, lift-off, and SEM imaging were carried out at the Purdue University Birck Nanotechnology Center.

FIGURE CAPTIONS

FIG. 1. (a) Nanoantenna geometry. Incident light polarized at 45° relative to the orthogonal symmetric (S) and antisymmetric (AS) axes results in the excitation of both antenna modes and a cross-polarized scattered field. (b) Two-dimensional contour plot showing the calculated light intensity before and after traversing the metasurface at $z = 0$. Illumination at 45° allows for spatial separation of the unperturbed and accelerating beams, which are readily observed at $z > 0$. (c) Calculated phase shift produced by the metasurface. The discontinuity in the phase gradient near $x = 135$ µm has a very minor effect on the beam properties.

FIG. 2. Schematics of the experimental setup and images of the metasurface. (a) Optical microscope image (green) showing the large-scale pattern of the antenna array, which modulates the incident field along the x-axis. Individual antennas are resolved in the SEM image. (b) Schematics describing the metasurface illumination and set-up used for imaging the accelerating beams. The beam reflected by the metasurface has been omitted. The first lens (L1) collects light from the accelerating beam scattered from defects in the glass and the second lens (L2) images the collected light onto the CCD. The gray area indicates the image plane.

FIG. 3. Experimental (a), (c) and calculated (b), (d) intensity plots for the 400 and 100 µm radius of curvature metasurfaces, respectively. To highlight the beam acceleration, a dotted line was added tracing circular trajectories of the same radius of curvature.

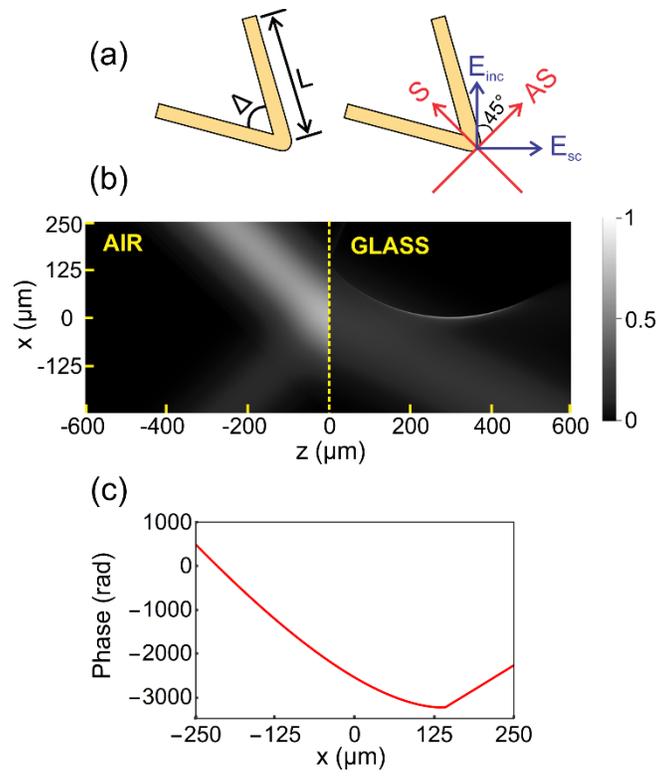

FIG. 1

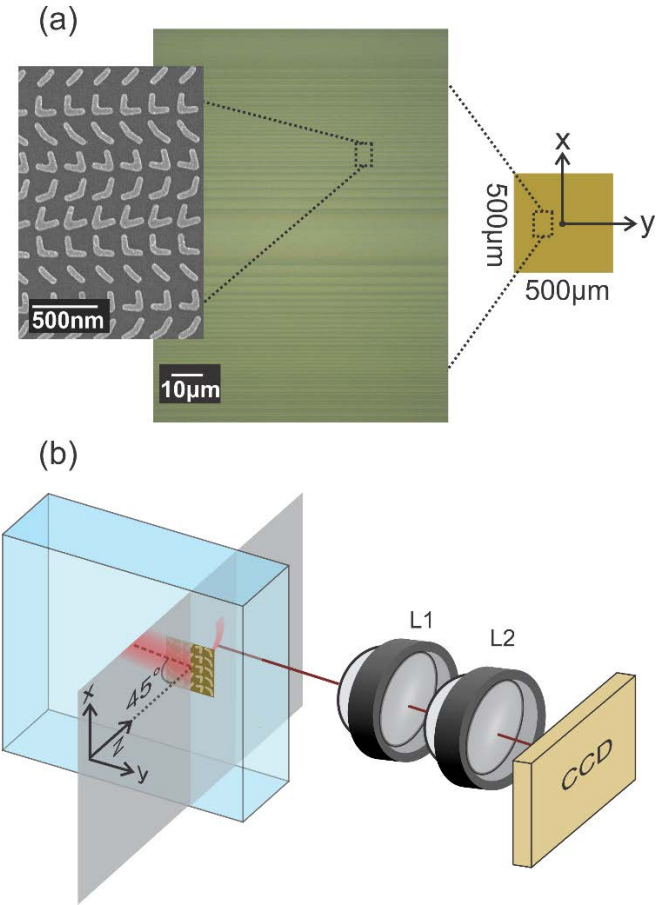

FIG. 2

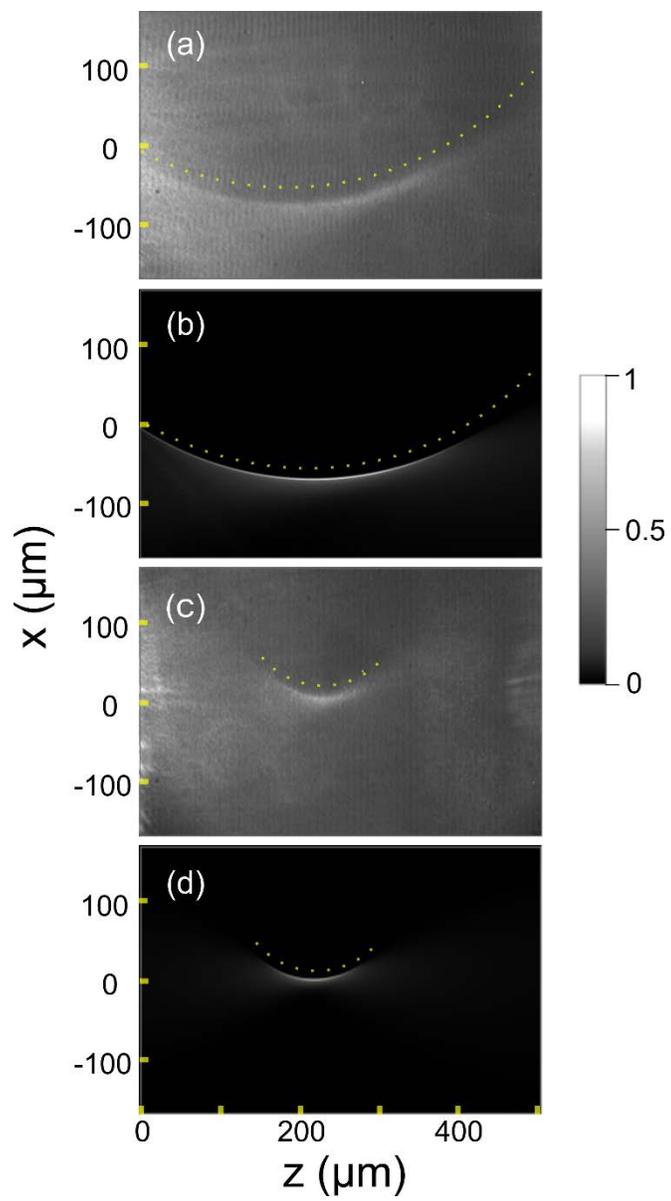

FIG. 3